\documentclass[twocolumn,PRB,aps,psfig,preprintnumbers,superscriptaddress]{revtex4}
\usepackage{graphicx}
\usepackage{epstopdf}
\usepackage{float}
\usepackage{color}
\usepackage{amsmath}

\begin{document}

\title{Slow Spin Dynamics in  the Hyper-Honeycomb Lattice  [(C$_2$H$_5$)$_3$NH]$_2$Cu$_2$(C$_2$O$_4$)$_3$ revealed by $^{1}$H NMR Studies}

\author{Q.-P. Ding}
\affiliation{Ames Laboratory, Iowa State University, Ames, Iowa 50011, USA}
\affiliation{Department of Physics and Astronomy, Iowa State University, Ames, Iowa 50011, USA}
\author{C. Dissanayake}
\affiliation{Department of Physics, University of Central Florida, Orlando, Florida 32816, USA}
\author{Santanu Pakhira}
\affiliation{Ames Laboratory, Iowa State University, Ames, Iowa 50011, USA}
\author{W.  J. Newsome}
\affiliation{Department of Chemistry, University of Central Florida, Orlando, Florida 32816, USA}
\author{ F. Uribe-Romo}
\affiliation{Department of Chemistry, University of Central Florida, Orlando, Florida 32816, USA}
\author{D. C. Johnston}
\affiliation{Ames Laboratory, Iowa State University, Ames, Iowa 50011, USA}
\affiliation{Department of Physics and Astronomy, Iowa State University, Ames, Iowa 50011, USA}
\author{Y. Nakajima}
\affiliation{Department of Physics, University of Central Florida, Orlando, Florida 32816, USA}
\author{Y. Furukawa}
\affiliation{Ames Laboratory, Iowa State University, Ames, Iowa 50011, USA}
\affiliation{Department of Physics and Astronomy, Iowa State University, Ames, Iowa 50011, USA}

\date{\today}

\begin{abstract} 
     We report the results of magnetic susceptibility $\chi$ and $^1$H nuclear magnetic resonance (NMR) measurements on a three-dimensional hyper-honeycomb lattice compound  [(C$_2$H$_5$)$_3$NH]$_2$Cu$_2$(C$_2$O$_4$)$_3$ (CCCO).
     The average value of the  antiferromagnetic (AFM) exchange coupling  between  the Cu$^{2+}$ ($S$ = 1/2) spins was determined to be $J$~$\sim$~50 K from the $\chi$ measurements. 
  No long-range magnetic ordering has been observed down to $T$~=~50~mK, although NMR lines become slightly broader at low temperatures below 1~K.
    The broadening of the NMR spectrum observed below 1 K reveals that the Cu spin moments remain at this temperature, suggesting  a non-spin-singlet ground state.      
The temperature and magnetic field dependence of 1/$T_1$ at temperatures above 20 K  is well explained by paramagnetic thermal spin fluctuations where the fluctuation frequency of Cu$^{2+}$ spins is higher than the NMR frequency of the order of MHz. 
     However, a clear signature of the slowing down of the Cu$^{2+}$ spin fluctuations was observed at low temperatures where 1/$T_1$ shows a thermally-activated behavior.
 The magnetic field dependence of  the magnitude of the spin excitation gap suggests that the magnetic behaviors of CCCO are characterized as an AFM chain at low temperatures.  


\end{abstract}

\maketitle


Magnetic frustration and quantum fluctuations, maximized for low spin $S$ = 1/2, are sources of a variety of fascinating phenomena  \cite{LB,PA,SS,CL,HTD}.
One of the exotic physical phenomena is a quantum spin-liquid state which breaks no symmetries down to zero temperature but exhibits macroscopic entanglement of strongly interacting spins and features exotic fractionalized excitations \cite{LB,PA}.
    Following  theoretical progress in understanding the quantum state, more and more candidate spin-liquid materials have been discovered and extensively
studied owing to advancements in material synthesis and experimental characterization techniques. 
    The most prominent spin-liquid candidates reported so far are $S$ = 1/2 kagom\'{e} lattices ZnCu$_{3}$(OH)$_{6}$Cl$_{2}$ \cite{TH2012,PM1_2011, JS2007,AO,TI,BF,PMRev, ImaiSci},  [NH$_{4}$]$_{2}$[C$_{7}$H$_{14}$N][V$_{7}$O$_{6}$F$_{18}$] \cite{LC}, \textit{S} = 1 hyperkagome Na$_{3}$Ir$_{4}$O$_{8}$ \cite{YO},  and $S$ = 1/2 triangular lattice organic compounds EtMe$_{3}$Sb[Pd(dmit)$_{2}$]$_{2}$ \cite{RK} and $\mathit{\kappa }$-(BEDT-TTF)$_{2}$Cu$_{2}$(CN)$_{3}$ \cite{YSZ,FLP,SY,SYQ}.

\begin{figure}[tb]
\includegraphics[width=8 cm]{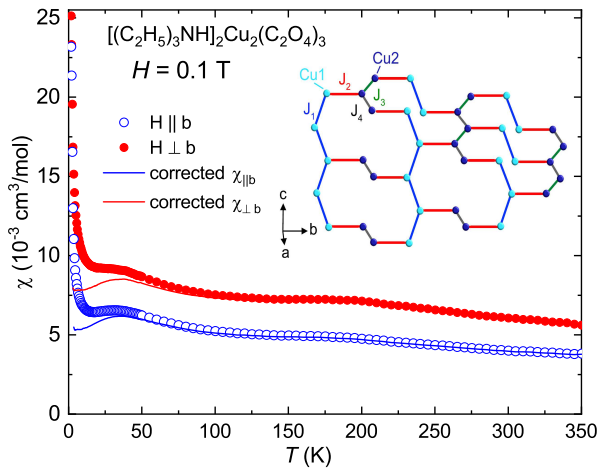} 
\caption{ $T$ dependence of the magnetic susceptibility $\chi$ at $H$ = 0.1 T for $H$ $||$ $b$ ($\chi_{|| b}$) and   $H$ $\perp$ $b$  ($\chi_{\perp b}$).  The solid lines are corrected $\chi$($T$) obtained by subtracting the respective impurity contributions.
    The inset shows the Cu$^{2+}$ network in CCCO where two Cu sites exist. Due to the Jahn-Teller distortion, magnetic interactions between Cu spins ($S$ = 1/2) are expected to be anisotropic: $J_{4}$ (Cu2-Cu2 with the shorter distance) $ > J_{1}$ (Cu1-Cu1) $ > J_{2}$ (Cu1-Cu2) $> J_{3}$  (Cu2-Cu2 with the longer distance) \cite{CuOx}.
}
\label{fig:Fig1}
\end{figure}

    Recently, a possible spin-liquid state has been reported in [(C$_2$H$_5$)$_3$NH]$_2$Cu$_2$(C$_2$O$_4$)$_3$ (hereafter, CCCO) \cite{CuOx} where Cu$^{2+}$ ($S$ =1/2) ions form a three-dimensional  hyper-honeycomb  lattice or  a ths net in the Reticular Chemistry Structure Resource  code notation \cite{RCSR}. 
    The schematic arrangement of Cu$^{2+}$ ($S$ = 1/2) is shown in the inset of Fig.~\ref{fig:Fig1} where the two different Cu sites (Cu1 and Cu2) form zigzag chains along the $c$ axis  and the $a$ axis direction, respectively,  and produce the four different magnetic interactions of $J_1$, $J_2$, $J_3$ and $J_4$ (for more detail, see supplementary material, SM \cite{SI}). 
     Although the magnetic interaction between Cu$^{2+}$ spins in the system seems to be complicated due to the structure and exchange paths,   the maximum antiferromagnetic (AFM) interaction of $J_4$  is reported to be $\sim$~200~K with  $J_{4}~ > ~J_{1} \sim 60~$K$~ >~J_{2} \sim 25~$K$~>~J_{3}$  in Ref.~\cite{CuOx}.
   Nevertheless, no sign of long-range magnetic-order down to a temperature $T$ = 60 mK has been observed by the $\mu$SR measurements \cite{CuOx}.
  The specific heat ($C$) measurement shows a relatively large value of $\gamma$ = 36 mJ/(Cu-mol$\cdot$K$^2$) for the linear term of $C$/$T$, indicating the gapless spin excitation in the ground state. 
Based on those experimental results, Zhang ${\it et~al.}$ \cite{CuOx}  suggested the ground state of  the compound is a gapless spin-liquid state.
     On the other hand, the quite recent  density-functional theory (DFT) calculations \cite{jacko21} suggested that the ground state of the compound may not be a spin-liquid state but can be regarded as  a weakly-coupled AFM Heisenberg chain system. 
      Despite the current interest in the compound as a spin-liquid candidate, the ground state of CCCO is still an open question,  and no detailed studies to characterize the  peculiar magnetic properties have yet been reported.

  In this paper,  we investigated the magnetic properties of the new $S$ = 1/2 hyper-honeycomb lattice compound CCCO by magnetic susceptibility and nuclear magnetic resonance (NMR) measurements. 
    NMR being a powerful local probe sheds light on the static and dynamic magnetic properties via spectra and nuclear spin-lattice relaxation rate (1/$T_1$) measurements, respectively, from a microscopic point of view. 
    From the $T$ dependence of the NMR spectrum,  we found  that the spin susceptibility does not vanish at 50 mK, suggesting a non-spin-singlet ground state in CCCO. 
    The $T$ dependence of $1/T_1$ indicates a slowing down of Cu spin fluctuations with decreasing $T$ down to 100 mK without showing any signature of long-range magnetic ordering.
     Our NMR data suggest that the magnetic behaviors of CCCO are characterized as an AFM chain system  where the Cu spins  fluctuate very slowly in the ground state.

    Rod-shaped single crystals of CCCO  were synthesized by the method described elsewhere \cite{zhang12a}.
    The size of each crystal is $\le1\times0.3 \times0.3$ mm$^3$ and we used several crystals which are aligned along the rod direction (the $b$ axis).
    Magnetic susceptibility $\chi$ measurements were carried out  using a Magnetic Property Measurement System (MPMS) from  Quantum Design, Inc., in the $T$ range 1.8--350 K under a magnetic field $H$ = 0.1 T for  two different magnetic field directions: $H$ parallel to the $b$ axis ($\chi_{||b}$) and perpendicular to the $b$ axis ($\chi_{\perp b}$). 
   NMR measurements down to 50 mK were performed  on $^{1}$H  ($I$ = 1/2, $\gamma_{\rm N}/2\pi$ = 42.5774 MHz/T) using a homemade phase-coherent spin-echo pulse spectrometer and an Oxford Kelvinox dilution refrigerator.  
    $H$ is applied perpendicular to the rod direction (i.e., $H$$\perp$$b$) for NMR measurements.
    $^1$H NMR spectra were obtained by Fourier transform of the Hahn spin-echo signals at fixed magnetic fields.
    The $^{1}$H spin-lattice relaxation rates (1/$T_{\rm 1}$) were measured with a saturation recovery method.
 The recoveries of longitudinal magnetization  display stretched exponential behavior due to distributions of $T_1$  values.       Therefore, $1/T_1$ at each $T$ is determined by fitting the nuclear magnetization $M$ versus time $t$ dependence after saturation using the stretched-exponential function $1-$$M$($t$)/$M({\infty})$ = exp[$-(t/T_{\rm 1})^{\rm \beta}$].    Here $M$($t$) and $M({\infty})$  are the nuclear magnetization at time $t$ after saturation and the equilibrium nuclear magnetization at time $t$ $\rightarrow$ $\infty$, respectively.   The $T$ and $H$ dependences of $\beta$ are shown in the Supplemental Material \cite{SI}. 
   The large distribution in $T_1$ seen in the reduction of the $\beta$ values is observed at low temperatures below $\sim$ 1 K where slow spin dynamics has been observed as discussed below. 


    Shown in Fig.~\ref{fig:Fig1}  is the $T$ dependence of the anisotropic $\chi_{||b}$ and  $\chi_{\perp b}$  in the $T$ range 1.8 $\leq $ \textit{T} $\leq $ 350~K. 
   $\chi(T)$ for both magnetic field directions exhibit broad humps around 170--190 K which has been attributed to a structural anomaly \cite{zhang12a,CuOx}. 
   The Curie-like increases in $\chi(T)$ at low temperatures are not intrinsic and  most likely originate from impurities,  as we do not see any corresponding effects on NMR spectrum described below. 
   The solid lines are corrected  $\chi(T)$ obtained by subtracting the impurity contributions.
   Broad maxima were observed at $T_{\rm max} \simeq$ 30--35 K,  one of the characteristic properties of  low-dimensional antiferromagnets originating from  short-range order of spins. 
   The average value of the Cu-Cu exchange interaction $J$ is estimated to be $J \sim k_{\rm B}T_{\rm max}/0.7 \simeq$~43--50 K using  a honeycomb lattice model  \cite{Johnston1997}. If we use an antiferromagnetic chain model as pointed out below, a similar value of $J \sim k_{\rm B}T_{\rm max}/0.641 \simeq$~47--54 K is obtained \cite{Johnston1997}, indicating the average value of $J$ can be consider to be of the order of $J$ $\sim$ 50 K. 
    The $T$ dependence of $\chi(T)$ of the single crystal is consistent with that of polycrystalline sample reported in Ref.~\cite{CuOx}.

\begin{figure}[tb]
  \includegraphics[width=\columnwidth]{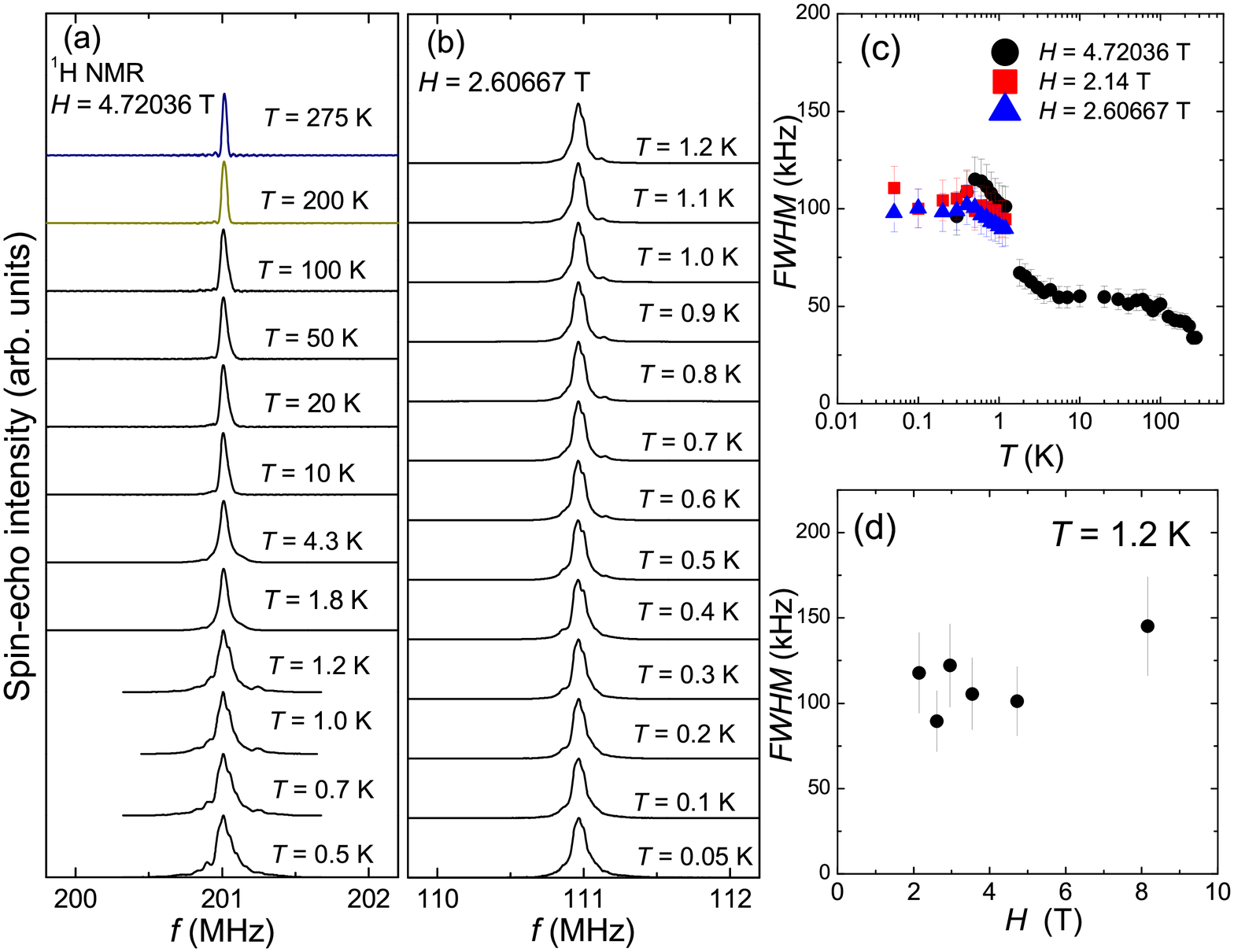}
	 \caption{(a) $T$ dependence of $^{1}$H  NMR spectra of CCCO at $H$ = 4.72036 T for the $T$ region 0.5--275 K. 
   (b) Same at $H$ = 2.60667 T for the low $T$ region 0.05--1.2 K.
   (c)  $T$ dependence of NMR line width (full width at half maximum, FWHM) of  $^{1}$H  NMR spectra.  
   (d) $H$ dependence of FWHM at 1.2 K. }
	\label{fig:Fig2}
\end{figure}

   Figures \ref{fig:Fig2}(a) and \ref{fig:Fig2}(b) show the typical $T$ dependencies of $^{1}$H-NMR spectra of  CCCO at $H$ = 4.72036~T for $T$~=~0.5--275~K  and $H$ = 2.60667 T for the low-$T$ region  $T$~=~0.05--1.2~K, respectively. 
   A sharp single NMR line with a full width at half maximum (FWHM) of 34 kHz was observed around zero NMR shift position at 275~K. 
   With decreasing temperature, spectra become slightly broader and, below 1.2 K a few shifted peaks with very small intensity were observed whose positions were nearly independent of $T$.
   Figure  \ref{fig:Fig2}(c) shows the $T$ dependence of the FWHM measured at  $H$ = 4.72036 T, together with the data at $H$ = 2.14 T and 2.60667 T.  
   With decreasing $T$, the FWHM increases slightly to $\sim$ 50 kHz down to 100 K and shows nearly $T$-independent behavior down to $\sim$ 4  K.
        In the present compound,  there are 32 protons in an ammonium cation, which would be located at slightly different distances from the magnetic ions and with different spatial orientations. 
   Therefore, a powder-pattern-like NMR spectrum will be observed due to the distributions of the internal field at the proton sites, even though we used single crystals.     
   In general, there are two possibilities for the origin of the internal field at the proton sites: classical dipolar  and contact-hyperfine fields from the Cu spins.
   The classical dipolar fields yield a broadening (and/or shoulder) of the line but no net shift, while a net shift of the line can be produced by the contact hyperfine field due to the overlap of the $s$-electron wave function of a proton with the $d$-electron wave function through the oxygen and nitrogen ions.
    No NMR line with a clear net shift is observed above 1.2 K. 
 This  indicates that the broadening of the line is mainly attributed to  the classical dipolar effects. 
    In this case, the FWHM of the NMR line is expected to be proportional to $\chi$. 
    Thus the increase of FWHM with decreasing $T$ from 300 K to 100 K  can be attributed to the increase of $\chi$.  
     The $T$-independent behavior of FWHM below 100 K could be due to the gradual structural transition around 150 K and also a disorder-order transition of one ethylene group on one of the ammonium cation at 165 K \cite{zhang12a,CuOx}, which produces the inhomogeneous broadening of the lines, masking the broad maximum around 35 K expected from the $T$ dependence of $\chi$.

      With further decreasing $T$, the FWHM increases up to $\sim$ 100 kHz and levels off below 1 K. 
     The FWHM at low $T$ is nearly independent of $H$ as shown in Fig. \ref{fig:Fig2}(d).
     The nearly $H$-independent behavior of FWHM at low $T$ indicates that the broadening of the spectrum arises from a nearly static internal field.
    The observation of a static internal field at nuclear sites depends on the fluctuation frequency of the spins. 
    Whenever the fluctuation frequency is lower than NMR frequency, one may observe static internal fields at nucleus sites. 
     On the other hand, the internal fields are time averaged to be zero when the fluctuation frequency is higher than the NMR frequency.
    The observation of the static internal field at the H sites therefore indicates that the fluctuation frequency of the Cu$^{2+}$ spins  is less than the NMR frequency of the order of MHz, evidencing the slowing down  of the Cu$^{2+}$  spins at low $T$, which is also consistent with the $T_1$ measurements described below.
    Thus, our experimental data clearly indicate that the Cu$^{2+}$ spin moments exist at the lowest $T$ and the ground state of the compound is not a spin-singlet state, consistent with a gapless ground state.


\begin{figure}[tb]
  \includegraphics[width=\columnwidth]{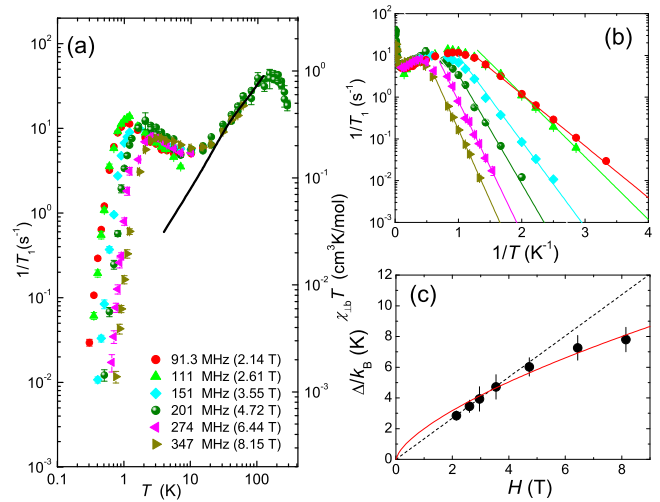}
	 \caption{(a) $T$ dependence of $^{1}$H  spin-lattice relaxation rates (1/$T_1$) at different frequencies. 
The solid line represents the $T$ dependence of $\chi_{\perp b} T$ (right scale). 
     (b) Semilog plot of 1/$T_1$ vs. 1/$T$. The solid lines are the best fits with the relation of 1/$T_1$ $\propto$ exp$(-\Delta/k_{\rm B}T)$ for each $H$.
    (c)  $H$ of  $\Delta/k_{\rm B}$  estimated from the slopes shown in  Fig. 3(b). Black dashed line: $\Delta/k_{\rm B}$ = 1.34$H$, red curve: $\Delta/k_{\rm B}$ = 2.0$H^{2/3}$.   
 }
	\label{fig:Fig3}
\end{figure}

     To investigate the dynamical properties of the Cu$^{2+}$ spins and the ground-state properties, we have performed proton spin-lattice relaxation rate 1/$T_{1}$ measurements in a wide $T$ range 0.1--275 K.
   Figure  \ref{fig:Fig3}(a) shows the $T$ dependence of $1/T_1$ of $^{1}$H NMR at various $H$.
  1/$T_1$ exhibits a peak around 150 K, which is probably related to the freezing of the rotational  motion of the C$_2$H$_5$ groups at the disorder-order transition at 165 K \cite{zhang12a,CuOx}.
   Below that temperature, 1/$T_1$ decreases and starts to increase at $\simeq$ 10 K, and then exhibits a peak at  $\simeq$ 1 K.
   It is noted that a nearly $H$ independent behavior of 1/$T_1$ is observed at high $T$ while 1/$T_1$ strongly depends on $H$ at low $T$, especially below $\sim$ 1 K.    
   As $H$ increases, the peak temperature of 1/$T_1$ shifts to higher $T$ and, at the same time, the height of the 1/$T_1$ peak becomes lower.
  Below $\sim$ 1 K, $T_1$ becomes longer with decreasing $T$ and the $T$ dependence is well reproduced by the thermal-activation behavior 1/$T_1$ $\propto$ exp$(-\Delta/k_{\rm B}T)$ as clearly seen in the semilogarithmic plot of 1/$T_1$ vs. $T$ in Fig. \ref{fig:Fig3}(b).
   Figure \ref{fig:Fig3}(c) shows the $H$ dependence of $\Delta/k_{\rm B}$ estimated from the slopes in  Fig.~\ref{fig:Fig3}(b).
   The values of $\Delta/k_{\rm B}$ seem to be proportional to $H$ and to follow the relation $\Delta/k_{\rm B}$ =  1.34$H$~K   (here $H$ is in units of Tesla) as shown by the dashed line up to $\sim$ 4 T, but the relation does not reproduce the $H$ dependence of $\Delta/k_{\rm B}$ above 5~T. 
   It is known that the magnitude of the field-induced gap is $\Delta/k_{\rm B}$ $\propto$ $H^{2/3}$ in AFM chains due to staggered $g$ tensors and/or  Dzyalonshinskii-Moriya (DM) interactions \cite{Oshikawa1997}.
  In fact, since there is no inversion symmetry at the middle point between the Cu1 and Cu2 sites (see the inset of Fig. 1), DM interaction can affect the magnetic properties of the system.
   As shown by the red curve, the $H$ dependence of $\Delta/k_{\rm B}$ is roughly reproduced by the relation.
    This would be consistent with the prediction of the recent DFT calculations suggesting a weakly-coupled AFM Heisenberg chain system \cite{jacko21}. 
   It should be emphasized that the magnitude of $\Delta/k_{\rm B}$ is estimated to be zero at $H$ = 0 from the $H$ dependence of  $\Delta/k_{\rm B}$ in Fig.~\ref{fig:Fig3}(c).
   This indicates no finite gap at zero magnetic field, in turn suggesting a gapless ground state in CCCO.
It is noted that, although we cannot exclude a possible finite critical field for the $H$-induced gap, this will not change our conclusion.

   It is important to point out that our $T_1$ data reveal a slowing  down of Cu$^{2+}$  spin fluctuations at low $T$. 
1/$T_1$ is generally expressed by the Fourier transform of the time correlation function of the transverse component $h_{\pm}$ of the fluctuating local field at nuclear sites with respect to the nuclear Larmor frequency $\omega_{\rm N}$ as \cite{Abragam},
 \begin{eqnarray}
\frac{1}{T_1} 
 =  \frac{1}{2} \gamma_{\rm N}^2 \int_{-\infty}^{+\infty} \langle \ h_{\pm}(t) h_{\mp}(0) \ \rangle 
{\rm exp}(i \omega_{\rm N}t) dt.  
 \label{eqn:$T_1$}
 \end{eqnarray}
  Since the internal field at the H sites is mainly due to classical dipolar field, 1/$T_1$ is given by a sum of two contributions   (1/$T_1$)$_{||}$ and  (1/$T_1$)$_{\perp}$ due to the magnetic fluctuations parallel ($||$) and perpendicular ($\perp$) to the external field, respectively \cite{Moriya1956}.  
Assuming the time correlation function decays as exp(-${\it \Gamma}$$t$), 1/$T_1$ can be written by \cite{Hone1974, Maegawa1995, Giovannini1971,T1comment}, 
 \begin{eqnarray}
\frac{1}{T_1}   =   A_{||}\chi_{||}T  \frac{{\it \Gamma}}{\it {\Gamma}^{\rm 2} + \omega_{\rm N}^{\rm 2}} + A_{\perp}\chi_{\perp}T  \frac{{\it \Gamma}}{\it {\Gamma}^{\rm 2} + \omega_{\rm e}^{\rm 2}},  
 \label{eqn:$T_1$_6}
\end{eqnarray}
where $\omega_{\rm e}$ is the electron Larmor frequency. 
 $A_i$ and $\chi_i$ are the parameters related to the local fields at the H sites and the magnetic susceptibilities, respectively, for the  $i$ (= $||$ and ${\perp}$) directions.   
Here the $\it {\Gamma}$ is the inverse of the correlation time of the fluctuation of the local fields at the H sites due to Cu$^{2+}$ spins and is assumed to be  isotropic for simplicity. 
 As we discuss below, since the first term of Eq.~\ref{eqn:$T_1$_6} is considered to be dominant, we consider only the first term in the following.

    When ${\it \Gamma}$  is independent of $T$, the $T$ dependence of 1/$T_1$ is simply proportional to $\chi_{||} T$. 
   The solid curve in Fig. \ref{fig:Fig3}(a) shows the temperature dependence of $\chi_{\perp b} T$ where we used $\chi_{\perp b}$ for $\chi_{||}$ since the magnetic field is applied perpendicular to the $b$ axis. 
   As can be seen in the figure, the $T$ dependence of 1/$T_1$ scales with that of $\chi_{\perp b} T$ above $\sim$20 K where 1/$T_1$ is nearly independent of $H$. 
   The result indicates that the nuclear relaxations are induced by the paramagnetic fluctuations of Cu$^{2+}$ spins whose frequency is much higher than the NMR frequency.

\begin{figure}[t]
  \includegraphics[width=\columnwidth]{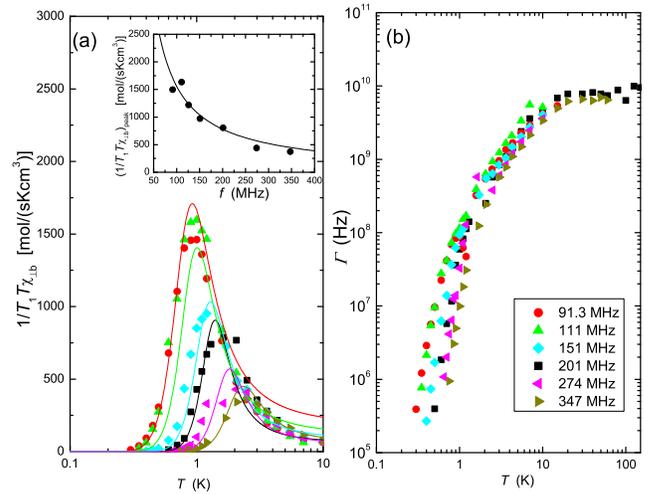}
	 \caption{(a) $T$ dependence of  1/${T_1T \chi_{\perp b}}$ measured at different resonance frequency. 
     Inset shows the resonance frequency dependence of the peak value of  1/${T_1T \chi_{\perp b}}$. 
     The solid line is the expected behavior based on Eq.~\ref{eqn:$T_1$_6}. 
    (b) $T$ dependence of ${\it \Gamma}$ estimated from $^{1}$H-1/$T_1$ data }
	\label{fig:Fig4}
\end{figure}

   Below $\sim$20 K, on the  other hand, the simple paramagnetic-fluctuation model cannot reproduce the experimental results. 
   To analyze the $T$ and $H$ dependencies of 1/$T_1$ at low $T$ using Eq.~\ref{eqn:$T_1$_6}, we re-plot the data by changing the vertical axis from 1/$T_1$ to 1/$T_1 \chi_{\perp b} T $  as shown in Fig. \ref{fig:Fig4}(a), where  $\chi_{\perp b}$ is assumed to be constant below 1 K since no clear change in the NMR shift and the FWHM were observed from the NMR spectra.  
  According to Eq.~\ref{eqn:$T_1$_6}, 1/${T_1T \chi_{\perp b}}$ is proportional to 1/${\it \Gamma}$  when ${\it \Gamma}$ $\gg$  $\omega_{\rm N}$  (fast-motion regime).
  This is actually observed above 20 K as discussed above. 
   On the other hand, 1/${T_1T \chi_{\perp b}}$  is proportional to ${\it \Gamma}$/$\omega_{\rm N}^2$ in the case of ${\it \Gamma}$ $\ll$  $\omega_{\rm N}$ (slow-motion regime) where 1/$T_1$ is expected to depend inversely on the square of the $H$ since $\omega_{\rm N}$ = $\gamma_{\rm N}H$. 
   When ${\it \Gamma}$ = $\omega_{\rm N}$, 1/${T_1T \chi_{\perp b}}$ reaches a maximum value.  
   
    Assuming $A$ = 1.9 $\times$ 10$^{12}$~rad$\cdot$mol/cm$^{3}$/K/s$^2$  and ${\it \Gamma}$ = ${\it \Gamma}_0$exp($-\Delta / k_{\rm B}T)$  where we used ${\it \Gamma}_0$ $\sim$ 1.0 $\times$ 10$^{10}$~Hz which slightly depends on $H$, the experimental results are well reproduced by Eq.~\ref{eqn:$T_1$_6}, as shown in Fig.~\ref{fig:Fig4}(a) by solid curves for different $H$. 
  The dependence of the peak height of 1/${T_1T \chi_{\perp b}}$ on the NMR resonance frequency (i.e., $H$) is also well reproduced by the model as shown by the solid curve in the inset of Fig.~\ref{fig:Fig4}(a).   
   These results indicate that the peak observed in the $T$ dependence of 1/${T_1T \chi_{\perp b}}$ originates from a crossover between the fast-motion and
the slow-motion regimes, whereby the fluctuation frequency of the Cu$^{2+}$ spins below the peak temperature is less than the NMR frequency range which is of the order of MHz.
    This is consistent with the observation of the nearly $H$-independent broadening of NMR spectrum  below 1 K. 
If we consider that the second term of Eq.~\ref{eqn:$T_1$_6} is relevant  for the peaks of 1/$T_1$, ${\it {\Gamma}}$ at the peak position is the electron Larmor frequency, much higher than NMR frequency, which will not produce the broadening of the NMR spectrum  observed below $\sim$1 K.
 These results suggest that 1/$T_1$ is mainly determined by the first term as pointed out, and thus we consider only the first term in our analysis.

     To derive the $T$ dependence of the fluctuation frequency for the Cu$^{2+}$ spins in a wide $T$ region, we extract the $T$ dependence of  ${\it \Gamma}$   from the 1/${T_1T \chi_{\perp b}}$ data,  assuming Eq.~\ref{eqn:$T_1$_6} is valid for all $T$ regions. 
    The estimated $T$ dependences of  ${\it \Gamma}$   for different $H$ are shown in Fig.~\ref{fig:Fig4}(b). 
    ${\it \Gamma}$ shows a thermally-activated behavior at low $T$ and is almost constant with ${\it \Gamma}$  $\sim$ 10$^{10}$ Hz at high $T$. 
    At $T$ below $\sim$ 1 K, the Cu$^{2+}$ spins fluctuate with low frequencies, which is less than the NMR frequency of the order of MHz.
    No loss of NMR signal intensity and the absence of critical slowing down rule out the possibility of a spin-glass phase down to 50 mK in CCCO. 
   This is further supported by the absence of a critical divergence of 1/$T_1$ or a cusp structure in 1/$T_1$ generally observed in a spin-frozen state \cite{Furukawa2015}.

In summary, we have carried out magnetic susceptibility and $^1$H NMR measurements in a wide $T$ range from 0.05 K to 350 K on the quantum spin-liquid candidate CCCO to investigate its magnetic properties, especially focusing on its ground-state magnetic properties.
  Although the average AFM exchange interaction is $J$~$\sim$~50~K from the broad maxima in the $T$ dependence of $\chi$,  no long-range magnetic ordering down to 50 mK has been observed by our NMR measurements, consistent with the $\mu$SR measurements \cite{CuOx}.  
 We found that the fluctuation frequency of the Cu$^{2+}$ spins slows down less than the NMR frequency of the order of MHz at $T \leq$ 1 K. 
    In addition, from the $T$ and $H$ dependence of 1/$T_1$, the application of $H$ is found to give rise to a spin-excitation gap whose $H$ dependence is well reproduced by $\Delta/k_{\rm B}$~$\propto$~$H^{2/3}$.     
    Our NMR data therefore suggest  that the spin dynamics of CCCO essentially is characterized as an AFM chain, not a spin-liquid state, as recently pointed out by DFT calculations \cite{jacko21}.

\appendix
The research was supported by the U.S. Department of Energy (DOE), Office of Basic Energy Sciences, Division of Materials Sciences and Engineering. Ames Laboratory is operated for the U.S. DOE by Iowa State University under Contract No.~DE-AC02-07CH11358. 
C.D. and Y.N. were supported by NSF CAREER DMR-1944975.


\begin{thebibliography}{10}
\bibitem{LB} L. Balents, Spin liquids in frustrated magnets, Nature \textbf{464}, 199 (2010) and references therein.
\bibitem{PA} P. A. Lee, An End to the Drought of Quantum Spin Liquids, Science \textbf{321,}1306 (2008).
\bibitem{SS} S. Sachdev, Quantum magnetism and criticality, Nat. Phys. \textbf{\ 4}, 173 (2008).
\bibitem{CL} C. Lacroix, P. Mendels, and F. Mila,\textit{\ Introduction to Frustrated Magnetism,} Springer Series in Solid-State Sciences (Springer, New York), Vol. 164.
\bibitem{HTD} \textit{Frustrated Spin Systems}, \textit{ed}. H. T. Diep (World Scientific, Singapore, 2005).

\bibitem{TH2012} T.-H. Han, J. S. Helton, S. Chu, D. G. Nocera, J. A. Rodriguez-Rivera, C. Broholm, and Y. S. Lee, Fractionalized excitations in the spin-liquid state of a kagome-lattice antiferromagnet, Nature \textbf{492}, 406 (2012).
\bibitem{PM1_2011} M. Jeong, F. Bert, P. Mendels, F. Duc, J.\thinspace C. Trombe, M.\thinspace A. de Vries, and A. Harrison, Field-Induced Freezing of a Quantum Spin Liquid on the Kagome Lattice, Phys. Rev. Lett. \textbf{107}, 237201 (2011).

\bibitem{JS2007} J.\thinspace S. Helton, H. Martinho, M.\thinspace S. Sercheli, P.\thinspace G. Pagliuso, D.\thinspace D. Jackson, M. Torelli, J.\thinspace W. Lynn, C. Rettori, Z. Fisk, and S.\thinspace B. Oseroff, Spin Dynamics of the Spin-1/2 Kagome Lattice Antiferromagnet ZnCu$_3$(OH)$_6$Cl$_2$, Phys. Rev. Lett.  \textbf{98}, 107204 (2007).

\bibitem{AO} A. Olariu, P. Mendels, F. Bert, F. Duc, J. Trombe, M. de Vries, and A. Harrison, $^{17}$O NMR Study of the Intrinsic Magnetic Susceptibility and Spin Dynamics of the Quantum Kagome Antiferromagnet ZnCu$_3$(OH)$_6$Cl$_2$, Phys. Rev. Lett. \textbf{100}, 087202 (2008).

\bibitem{TI} T. Imai, E. A. Nytko, B. M. Bartlett, M. P. Shores, and D. G. Nocera, $^{65}$Cu, $^{35}$Cl, and $^{1}$H NMR in the $S$ = $\frac{1}{2}$ Kagome Lattice ZnCu$_3$(OH)$_6$Cl$_2$, Phys. Rev. Lett. \textbf{100}, 077203 (2008).

\bibitem{BF} B. F\aa k, E. Kermarrec, L. Messio, B. Bernu, C. Lhuillier, F. Bert, P. Mendels, B. Koteswararao, F. Bouquet, J. Ollivier, A. D. Hillier, A. Amato, R. H. Colman, and A. S. Wills, Kapellasite: A Kagome Quantum Spin Liquid with Competing Interactions, Phys. Rev. Lett. \textbf{109}, 037208 (2012). 


\bibitem{PMRev} P. Mendels and F. Bert, Quantum Kagome Antiferromagnet ZnCu$_3$(OH)$_6$Cl$_2$, J. Phys. Soc. Jpn. \textbf{79}, 011001 (2010). 

\bibitem{ImaiSci} H. Fu, T. Imai, T.H. Han, Y. S. Lee, Evidence for a gapped spin-liquid ground state in a kagome Heisenberg antiferromagnet, Science {\bf 350}, 655 (2015). 
 

\bibitem{LC} L. Clark, J. C. Orain, F. Bert, M. A. De Vries, F. H. Aidoudi, R. E. Morris, P. Lightfoot, J. S. Lord, M. T. F. Telling, P. Bonville, J. P. Attfield, P. Mendels, and A. Harrison, Gapless Spin Liquid Ground State in the $S$~= 1/2 Vanadium Oxyfluoride Kagome Antiferromagnet [NH$_4$]$_2$[C$_7$H$_{14}$N][V$_7$O$_6$F$_{18}$], Phys. Rev. Lett. \textbf{110}, 207208 (2013). 


\bibitem{YO} Y. Okamoto, M. Nohara, H. Aruga-Katori, and H. Takagi, Spin-Liquid State in the $S$~= 1/2 Hyperkagome Antiferromagnet Na$_4$Ir$_3$O$_8$, Phys. Rev. Lett. \textbf{99}, 137207 (2007). 



\bibitem{RK} T. Itou, A. Oyamada, S. Maegawa, and R. Kato, Instability of a quantum spin liquid in an organic triangular-lattice antiferromagnet, Nat. Phys.  \textbf{6}, 673 (2010). 


\bibitem{YSZ} Y. Shimizu, K. Miyagawa, K. Kanoda, M. Maesato, and G. Saito, Spin Liquid State in an Organic Mott Insulator with a Triangular Lattice, Phys. Rev. Lett. \textbf{91}, 107001 (2003). 

\bibitem{FLP} F. L. Pratt, P. J. Baker, S. J. Blundell, T. Lancaster, S. Ohira-Kawamura, C. Baines, Y. Shimizu, K. Kanoda, I. Watanabe and G. Saito, Magnetic and non-magnetic phases of a quantum spin liquid, Nature \textbf{471},612 (2011). 

\bibitem{SY} S. Yamashita, Y. Nakazawa, M. Oguni, Y. Oshima, H. Nojiri, Y. Shimizu, K. Miyagawa, and K. Kanoda, Thermodynamic properties of a spin-1/2 spin-liquid state in a $\kappa$-type organic salt, Nat. Phys. \textbf{4}, 459 (2008). 


\bibitem{SYQ} Y. Qi, C. Xu, and S. Sachdev, Dynamics and Transport of the $Z_2$ Spin Liquid: Application to $\kappa$-(ET)$_2$Cu$_2$(CN)$_3$, Phys. Rev. Lett. \textbf{102}, 176401(2009). 

\bibitem{CuOx}  B. Zhang, P. J. Baker, Y. Zhang, D. Wang, Z. Wang, S. Su, D. Zhu, and F. L. Pratt, Quantum Spin Liquid from a Three-Dimensional Copper-Oxalate Framework, J. Am. Chem. Soc. {\bf 140}, 122 (2018).


\bibitem{RCSR} M. O\rq{}Keeffe, M. A. Peskov, S. J. Ramsden, and O. M. Yaghi, The Reticular Chemistry Structure Resource (RCSR) Database of, and Symbols for, Crystal Nets, Acc. Chem. Res., {\bf41},  1782  (2008).



\bibitem{SI}  See supplemental material for the experimental details, the crystal structure and a stretched-exponential fitting for the measured nuclear magnetization recovery behavior. 

\bibitem{jacko21} A. C. Jacko and B. J. Powell, Quasi-one dimensional magnetic interactions in the three-dimensional hyper-honeycomb framework [(C$_2$H$_5$)$_3$NH]$_2$Cu$_2$(C$_2$O$_4$)$_3$, Phys. Chem. Chem. Phys. {\bf  23}, 5012 (2021).

\bibitem{zhang12a} B. Zhang, Y. Zhang, and D. Zhu, [(C$_2$H$_5$)$_3$NH]$_2$Cu$_2$(C$_2$O$_4$)$_3$: a three-dimensional metal–oxalato framework showing structurally related dielectric and magnetic transitions at around 165~K, Dalton Trans. {\bf 41}, 8509 (2012).

\bibitem{Johnston1997} D. C. Johnston, Normal-state magnetic properties of single layer cuprate high-temperature  superconductors and related materials, Handbook of Magnetic Materials, {\bf 10}, 1 (1997).






\bibitem{Oshikawa1997} M. Oshikawa and I. Affleck, Field-Induced Gap in $S$ = 1/2 Antiferromagnetic Chains, Phys. Rev. Lett. {\bf 79}, 2883 (1997).

\bibitem{Abragam} A. Abragam, $The~Principles~of~Nuclear~Magnetism$ (Clarendon Press, Oxford, 1961).

\bibitem{Moriya1956} T. Moriya, Nuclear Magnetic Relaxation in Antiferromagnets, Prog. Theo. Phys. 16, 23 (1956).

\bibitem{Hone1974} D. Hone, C. Scherer, and F. Borsa, Proton spin-lattice relaxation in TMMC [(CH$_3$)$_4$NMnCl$_3$3], Phys. Rev. B. {\bf 9}, 965 (1974).


\bibitem{Maegawa1995} S.  Maegawa, Nuclear magnetic relaxation and electron-spin fluctuation in a triangular-lattice Heisenberg antiferromagnet CsNiBr$_3$,  Phys. Rev. B. {\bf 51}, 15979 (1995).


\bibitem{Giovannini1971} B. Giovannini P. Pincus,  G. Gladstone, and A. J. Heeger, Nuclear relaxation in dilute magnetic alloys, J. Phys. Colloques, {\bf 32} (C1) 163 (1971).  

\bibitem{T1comment} In Refs. \cite{Hone1974,Maegawa1995}, $\omega_{\rm N}$ was set to zero. But, here we do not set  $\omega_{\rm N}$ = 0  as  in the case of Ref.  \cite{Giovannini1971}. 




\bibitem{Furukawa2015}  T. Furukawa, K. Miyagawa, T. Itou, M. Ito, H. Taniguchi, M. Saito, S. Iguchi, T. Sasaki, and K. Kanoda, Quantum Spin Liquid Emerging from Antiferromagnetic Order by Introducing Disorder,  Phys. Rev. Lett. {\bf 115}, 077001 (2015).

 \end{thebibliography}
\end{document}